# Adiabatic Invariants in Stellar Dynamics:
# III. Application to Globular Cluster Evolution[1]

Martin D. Weinberg[2]
Department of Physics and Astronomy
University of Massachusetts/Amherst

## ABSTRACT

The previous two companion papers demonstrate that slowly varying perturbations do not result in adiabatic cutoffs and provide a formalism for computing the long-term effects of time-dependent perturbations on stellar systems. Here, the theory is implemented in a Fokker-Planck code and a suite of runs illustrating the effects of shock heating on globular cluster evolution are described.

Shock heating alone results in considerable mass loss for clusters with $R_g \lesssim 8\,\mathrm{kpc}$: a concentration $c = 1.5$ cluster with $R_g = 8\,\mathrm{kpc}$ loses up to 95% of its initial mass in 15 Gyr. Only those with concentration $c \lesssim 1.3$ survive disk shocks inside of this radius. Other effects, such as mass loss by stellar evolution, will increase this survival bound. Loss of the initial halo together with mass segregation leads to mass spectral indices, $x$, which may be considerably larger than their initial values.

## 1. Introduction

A globular cluster is stripped by the differential acceleration and heated by its time-dependent acceleration through the Galaxy. Both of these external effects decrease the central concentration and cause the cluster to be less bound. This trend is countered by two-body relaxation which increases its central concentration. Many clusters, especially those within the solar circle, would be disrupted without two-body relaxation. Understanding the interplay of this competition is crucial to understanding the evolutionary history of the overall globular cluster population.

---

[1]A PostScript file of this paper including figures can be obtained by anonymous FTP (ptolemy.phast.umass.edu:pub/shock3.ps.Z)

[2]Alfred P. Sloan Foundation Fellow

This interplay has been studied with a wide variety of approximations and techniques. The two-body relaxation is straightforwardly followed using Fokker-Planck techniques with tidal stripping is traditionally included as a boundary condition (e.g. Cohn 1980, Lee & Ostriker 1987, Chernoff & Weinberg 1990, CW). Oh and Lin (1992) have investigated the simultaneous interaction of relaxation and the tidal field with a more realistic hybrid approach. Chernoff et al.(1986, CKS, see also Chernoff and Shapiro 1987) and Aguilar et al.(1988) have tried to include all three effects by mapping the evolution along sequences of King models and by Monte Carlo simulation, respectively.

This paper focuses on the effect of gravitational shocking due to disk interactions, perhaps the least well-explored of the external effects. Previous work treats shocking using the impulse approximation with an adiabatic cutoff. The current investigation is motivated by the previous two papers in this series (Papers I and II). Together they show that the classical adiabatic criterion does not hold for general stellar systems and may lead to significant underestimate for this time-dependent heating.

To summarize, the standard adiabatic criterion is based on the harmonic oscillator model. If the oscillator is subjected to a slowly varying perturbation, the change in action is exponentially small, $\exp(-\Omega/\nu)$ where $\nu$ is the characteristic frequency for the slow change and $\Omega$ is the oscillator frequency. In a multidimensional system, such as a stellar system, each independent degree of freedom has an independent frequency of oscillation. As long as the frequency of the perturbation is small compared any linear combination of these frequencies, each combination of frequencies acts like a single oscillator and constants of motion for the original orbit will be conserved. However, if one of those combinations is zero or nearly so then viewed in the appropriate frame of reference, the perturbation is fast compared to the orbit and the orbit gets "kicked." This is similar to a resonance but with arbitrarily small resonant frequency. The basic mechanism is discussed in detail in Paper I. Since a realistic stellar system has frequencies continuously distributed in some range, there will almost always be some low-order commensurabilities and the corresponding orbits will *not* be adiabatically invariant. Averaging over the whole stellar system, the orbits with broken invariants can give an appreciable overall similar in magnitude to the impulsive contribution as shown in Paper II.

Paper II also provides a formalism and computational approach to computing the effects of a gravitational shock for *any* encounter rate. Specifically, the long-term change to the phase-space distribution function, $\langle f_2 \rangle$ (e.g. eqn. 30 of Paper II), is easily incorporated in to a Fokker-Planck calculation (§2). Examples (§3) are chosen to illustrate the physical and observational consequences of the new gravitational shocking theory and relaxation alone and do not represent the whole panoply of cluster physics. Nonetheless, this pilot study predicts that gravitational shocking may play a major role in cluster evolution for clusters with orbital radii inside the solar circle. Detailed consequences of shocking will be described in detail and incorporated into a wide-ranging grid of models in a future paper.

## 2. Fokker Planck implementation

### 2.1. Method

The one-dimensional Fokker-Planck models describe the evolution of a distribution of orbits with energy in time driven by two-body encounters. The angular momentum distribution is constrained to remain isotropic everywhere. Passage through the disk changes the orbit distribution in a fixed potential as described by equation (30) in Paper II:

$$\langle f_2 \rangle = \frac{V_o^2}{4h^4} \frac{\pi}{2(v_z/h)^2} \sum_{l=-\infty}^{\infty} \delta_{l_3 0} \left(\frac{1}{15} + \frac{43}{90}\delta_{l_2 0}\right) \left(\mathbf{l} \cdot \mathbf{\Omega} \frac{\partial}{\partial E} + l_2 \frac{\partial}{\partial J}\right) \Big\{$$
$$e^{-(\mathbf{l}\cdot\mathbf{\Omega})^2/2(v_z/h)^2} \left(\mathbf{l} \cdot \mathbf{\Omega} \frac{df_o}{dE} + l_2 \frac{\partial}{\partial J}\right) \left|X_{l_2}^{l_1}\right|^2 \Big\}. \qquad (1)$$

This expression assumes that the disk has Gaussian profile, $\rho(z) = \rho_o \exp(-z^2/h^2)$ with $V_o \equiv 4\pi G \rho_o h^2$, and that the perpendicular component of cluster's orbital velocity relative to the disk is $v_z$. Calculations suggest that the results are insensitive to the details of the vertical profile. The quantity **l** is a three-vector of integers $l_j$, one for each degree of freedom, $\mathbf{\Omega}$ is the three-vector of frequencies and $X_{l_2}^{l_1}$ is proportional to the Fourier coefficient for the action-angle expansion of $r^2$ (see Paper II for details). Equation (1) may be averaged over all angular momenta at fixed energy to yield the perturbed isotropic distribution function:

$$\langle\langle f_2 \rangle\rangle(E) = \frac{\int dJ J/\Omega_1(E,J) \, \langle f_2 \rangle(E,J)}{\int dJ J/\Omega_1(E,J)}. \qquad (2)$$

Because $\langle\langle f_2 \rangle\rangle$ derives from a linearized solution, $f_o + \langle\langle f_2 \rangle\rangle$ may be less than zero near the edge of the original model; in this case, the new distribution is set to zero, e.g.:

$$f_{new} = \max(f_o + \langle\langle f_2 \rangle\rangle, 0). \qquad (3)$$

The energy at which $f_{new} = 0$ determines the new edge, $E_{edge} \equiv \Phi(r_{edge})$ where $\Phi$ is the gravitational potential for the background model. In addition to loss by shock heating, there is a dynamical tidal boundary at the effective x-point (e.g. CW). Ideally, both boundaries need to be implemented, but to isolate effects, only gravitational shocking is included here.

In the Fokker-Planck calculations described below, equations (2) and (3) are used with equation (1) to determine the shocking explicitly[3]. However, for some implementations, this approach may be inefficient and cumbersome. In the numerical examples below, the shocking calculations take up to 30% of the total CPU time. Instead, the perturbed distribution functions may be tabulated for a variety of concentrations. The perturbed distribution functions and their ratios with the unperturbed distributions are shown in Figures 1 and 2. Since the core radius is easily computed and the edge and potential are known, grids similar to those shown in these figures may be scaled and interpolated to compute the shocking.

Finally as a check, Figure 3 compares the predicted changes in the distribution due to a single shock using equation (2) with the results of restricted n-body simulation for a $W_0 = 5$ King model; the agreement is good. Since the change in the distribution is computed at fixed energy, stars with slightly larger energies than $E_{edge}$ will still be bound to the cluster leading to an overestimate of the change very near the edge.

## 2.2. Comparison to existing estimates

The total change in kinetic energy for a shock is

$$\Delta T = \int dE\, E\, p(E) f_{new}(E) \tag{4}$$

where $f_{new}$ is given by equation (3) and $p(E)$ is the usual phase space factor defining the differential number density $dN = f(E)p(E)dE$.

This expression for $\Delta T$ has been compared with the prescription given in CKS for isotropic King models with dimensionless potential in the range $1 < W_o < 9$. For their parameters, the overall heating from equation (4) is approximately a factor of three more than the impulse approximation with an adiabatic cutoff for orbits with 90° inclination. This factor largely is due to the more precise treatment, not excess heating in the adiabatic regime. Nonetheless, the adiabatic contribution for clusters with lower inclination orbits will be even larger than the roughly 10% seen in Figures 4–6 for higher inclination orbits. More importantly, the adiabatic heating perturbs orbits of much lower energy, causing significant evolution of the cluster (see §4).

The contribution to the shock heating in the adiabatic and impulsive regimes is estimated by separately summing the contribution to equation (2) for $\Omega/\nu$ greater than less than one, respectively. The quantity $\Omega \equiv \Omega_1 + \Omega_2$ for $J/J_{max}(E) = 1/\sqrt{2}$. This is shown in Figures 4–6 for models of three different concentrations. For the $W_0 = 3, 5, 7$ models, the adiabatic contribution is roughly 5%, 10%, 20% at 90° inclination (perpendicular to the disk). The adiabatic contribution increases with decreasing inclination since $v_z(i) = v_z(90°)\sin(i)$. For example, the fractional for the $W_0 = 5$ model for a 45° is roughly 30% and increases sharply for smaller inclinations. The frequency at peak heating (negative $\Delta T$) is dominated by the adiabatic contribution. Note that the cluster is "cooled" (positive for $\Delta T$) small frequencies in Figure 4. In this case, the halo is redistributed by the shock but because of the low binding energies, only those scattered to higher binding energies remain bound to the cluster, leading to a net decrease in total energy.

---

[3]See the Appendix of Paper II for a possible computational strategy.

## 3. Model parameters

To investigate the effects of disk shocking on cluster evolution, we apply this shocking scheme to clusters with the following features:

1. One-decade mass spectrum, $m \in [0.1, 1.0]\,\mathrm{M}_\odot$, to represent the stellar population at the current epoch.

2. Power law mass function with initially everywhere constant mass spectral index $x$.

3. An exponential Galactic disk, $\Sigma = \Sigma_o e^{-R/3.5\,\mathrm{kpc}}$ following the Bahcall & Soneira model (1980), which controls the amplitude of the gravitational shocking.

4. An Gaussian profile for the disk, $\rho(z) = \rho_o e^{-z^2/(325\,\mathrm{pc})^2}$, with a passage frequency $\nu$ determined by a constant orbital velocity, $\nu = V_{orb}/h$, $V_{orb} = 200\,\mathrm{km\,s^{-1}}$.

5. Initially isotropic King models with a range of concentrations, $0.6 < c < 1.5$ and galactocentric radii, $3\,\mathrm{kpc} \leq R \leq 16\,\mathrm{kpc}$, and $M = 3 \times 10^5\,\mathrm{M}_\odot$.

The models will be run to either core collapse or until disruption. Core collapse is simply used as a convenient stopping point in this preliminary investigation. A later set of runs will follow evolution through core collapse and a variety of environmental effects. In the spirit of both testing the importance of tidal shocking alone and its relevance to present-day clusters, these runs include no mass loss by stellar evolution.

## 4. Results

### 4.1. Description of the evolution

Figure 7 compares a cluster at $R = 8\,\mathrm{kpc}$ and initial concentration $c = 1.5$ without shocking, with shocking but including an adiabatic cutoff and with shocking using equations (1)–(3). In all cases the models reach core collapse. In the unshocked case the mass at the final time is $M = 0.98$ compared to $M = 0.15, 0.05$ for the two shocked cases. Clearly, the gravitational shocking strengthens the expansion and subsequent mass loss. Although the evolution with shocking is dramatic in both cases, the full theory results in a cluster which is three times less massive and 95% truncated. Time to core collapse is shorter in the shocked case due to the shorter local relaxation times in the less massive clusters (but this is not true for all cases, see below).

Figure 8 compares three clusters at $R_g = 12\,\mathrm{kpc}$, all with shocking, and concentrations $c = 1.5, 1.0$, and $0.67$. The two least concentrated eventually disrupt but only after $2.5 \times 10^{10}\,\mathrm{yr}$ and $5 \times 10^{10}\,\mathrm{yr}$. The most concentrated model does core collapse but more slowly in this case than without shocking due to the heating and mass loss which expands the cluster and balances the tendency for relaxation to contract the core. Notice that the disk shocking alone has effectively halted the core-collapse. The middle panel shows that shock heating may balance relaxation. The inner Lagrangian radii of the cluster are nearly constant, although the outer cluster is continuously stripped.

### 4.2. Effects of cluster orbit and concentration

Figure 9 summarizes the end states for the trial grid of runs ranging in galactocentric radius and initial concentration. As expected, the mass loss is more extensive in the inner Galaxy, where the gravitational shock is stronger due to the increase in disk surface density. As the inclination decreases, the shocking increases in strength by a factor of 2 before decreasing again for low inclinations. (cf. Figs. 4–6). For very low inclinations, the cluster should really be treated as a member of the disk and the model used here will be incorrect. Nonetheless, the time-dependent forcing by the disk on the cluster as it oscillates in the plane may be treated with the same formalism and will be reported later. Cluster parameters from Webbink (1985) are plotted in Figure 10 the $R_g$ vs. concentration plane along with the disruption boundary from Figure 9. We predict that the 15 clusters at or below the boundary are in the process of disrupting.

Figure 11 compares the edge computed by the criteria given in equation (3) and the tidal boundary used in CW for clusters of varying $R_g$. The theory predicts that the truncation should be at roughly 20% the inferred tidal boundary. A simultaneous treatment of both shocking and tidal stripping is necessary.

### 4.3. Time evolution of mass function

Previous researchers concluded that the mass function power-law index in the outer parts of the cluster would reflect the primordial values, even after the many central relaxation times expected in 15 Gyr (e.g. CW). Recently Richer et al.(1991) found that some clusters have very steep mass functions with $x$ between 2.5 and 3. The theory of shocking presented here predicts that the stars forming the present-day halo were inside the half-mass radius at early times and expanded after mass loss and heating due to strong gravitational shocking. This is illustrated in Figures 12 and 13 which show the mass spectral index $x$ as a function of time and relative position in the star cluster. The model in Figure 12 has no shocking and very little overall mass loss. The halo population—particles with relative radii $r/r_{edge} \gtrsim 0.2$—remains close to the original $x = 3$. Near $r/r_{edge} = 0.07$, $x$ increases due to mass segregation. The model in Figure 13 includes shocking and about 84% of its mass over before core collapse in $3.2 \times 10^{10}$ yr. In this case, parts of the cluster which suffer mass segregation inside of the half-mass radius at early times become the halo at the present epoch and beyond, producing an enhanced power-law index at $r/r_{edge} \lesssim 1.0$

### 5. Summary

The main conclusion of this paper are:

1. The gravitational shocking theory presented in Paper II predicts an overall heating rate larger by at least a factor of 2 over previous results (roughly a factor of 3 for King profiles) for orbits with 90° inclination (passage perpendicular to the disk). However, the peak heating is a factor of 2 larger still and occurs in the adiabatic regime. Moreover, heating in the adiabatic regime occurs at smaller binding energies than previously predicted which leads to significant dynamical evolution.

2. This enhanced heating *alone* may disrupt clusters with $c \lesssim 1.3$ and $R_g \lesssim 8\,\text{kpc}$ and delay core collapse for clusters with $c \gtrsim 1.3$ and $R_g \gtrsim 5\,\text{kpc}$ beyond the Hubble time. All in all, these results suggest that shock heating will play a defining role in the evolution of clusters inside the solar circle.

3. The present day cluster halos may have expanded from inside the half-mass radius and at early epoch due to efficient stripping of cluster halos by shock heating. Therefore, contrary to earlier claims (e.g. Chernoff and Weinberg 1990), the outer parts of clusters may NOT reflect the initial mass function. A cluster will initial power-law mass spectral index $x = 3$ may have $x \gtrsim 4$ in its halo at the current time.

The quantitative results quoted in §4 should only be used as relative indicators in the context of this paper since only restricted set of physical effects and initial conditions have been investigated. A full set of runs (in progress) will include shocking from both the disk and halo/spheroid for eccentric orbits as well as stellar evolution and binary interactions. The thick-disk clusters require special treatment since the vertical oscillation is comparable to or slower than the stellar periods and the disk passage is quasiperiodic rather than a single distinct shock; nonetheless, this case is straightforwardly treated using this theory and will be reported in a later paper.

I thank David Chernoff, Greg Fahlman, Chigurupati Murali, Doug Richstone and Scott Tremaine for stimulating discussions, and the Institute for Theoretical Physics in Santa Barbara for its hospitality. This work was supported in part by NSF grant PHY89-04035 to ITP and NASA grant NAGW-2224.

## REFERENCES

Aguilar, L., Hut, P., and Ostriker, J. P. 1988, Astrophys. J., 335, 720.

Bahcall, J. N. and Soniera, R. M. 1980, Astrophys. J., Suppl. Ser., 44(1), 73, Abstr. in 240, 374.

Chernoff, D. F., Kochanek, C. S., and Shapiro, S. L. 1986, Astrophys. J., 309, 183 (CKS).

Chernoff, D. F. and Shapiro, S. L. 1987, Astrophys. J., 322(1, part 1), 113.

Chernoff, D. F. and Weinberg, M. D. 1990, Astrophys. J., 351, 121 (CW).

Cohn, H. 1980, Astrophys. J., 242(2, part 1), 765.

Lee, H. M. and Ostriker, J. P. 1987, Astrophys. J., 322(1, part 1), 123.

Oh, K. S. and Lin, D. N. C. 1992, Astrophys. J., 386, 519.

Richer, H. B., Fahlman, G. G., Buonanno, R., Fusi-Pecci, F., Searle, L., and Thompson, I. B. 1991, Astrophys. J., 381, 147.

Webbink, R. F. 1985, in P. Hut and J. Goodman (eds.), *Dynamics of star clusters*, pp 541–577, D. Reidel Publishing Co., Dordrecht, Proceedings of the Symposium, Princeton, NJ, May 29-June 1, 1984.

Weinberg, M. D. 1994a, *Adiabatic Invariants in Stellar Dynamics: I. Basic concepts*, submitted (Paper I).

Weinberg, M. D. 1994b, *Adiabatic Invariants in Stellar Dynamics: II. Gravitational shocking*, submitted (Paper II).

FIGURE CAPTIONS

Fig. 1.— Perturbed distribution function as a function of scaled energy ($E \in [-1, 0]$) for King models with various $W_0$. The scaled energy equals $-1$ at the center and $0$ at the edge.

Fig. 2.— As in Fig. 1 but showing the ratio of the perturbed to unperturbed distribution function.

Fig. 3.— Comparison of direct simulation to perturbation theory calculation. Open circles show the change in particle number per energy bin due to a disk shock, $N - N_o$, computed from a direct integration in a fixed potential with $2 \times 10^5$ particles realized from a $W_0 = 5$ King model. The solid curve shows the predicted relation using eq. 2.

Fig. 4.— Top panel shows the heating for a $W_0 = 3$ King model cluster for a single disk shock as a function of passage frequency. Units are chosen so that total energy is $-1/8$. A cluster with $R_g = 8\,\mathrm{kpc}$, $v_z = 200\,\mathrm{km\,s^{-1}}$, and $h = 325\,\mathrm{pc}$ has $v_z/h = 1.1$. The solid curve shows the total heating, $\Delta T$, and the dotted and dashed curve shows the contribution in the adiabatic and impulsive regimes, respectively. The bottom panel shows the ratio of $\Delta T$ in the adiabatic and impulsive regimes.

Fig. 5.— Same as Fig. 4 but for $W_0 = 5$. The same orbit has the value $v_z/h = 0.67$.

Fig. 6.— Same as Fig. 4 but for $W_0 = 7$. The same orbit has the value $v_z/h = 0.33$.

Fig. 7.— Lagrangian radii for a cluster with $c = 1.5$ initially and $R = 8\,\mathrm{kpc}$. The evolution at the left has no gravitational shocking, the central panel includes shocking in the impulsive limit only, and the one at the right includes the full shocking calculation. The percent of mass enclosed is 2, 5, 10, 20, 30, 40, 50, 60, 70, 80, 90, 95, 98 from bottom to top. Time is shown in years. The fractional mass remaining at core collapse is 0.98, 0.15, and 0.055, respectively.

Fig. 8.— As in Fig. 7 but all with gravitational shocking, $R = 12\,\mathrm{kpc}$ and concentration $c = 1.5, 1.0, 0.67$ from left to right.

Fig. 9.— Summary of runs for $V_{orb} = 200 \,\mathrm{km\,s^{-1}}$. Points locate values of galactocentric radius $R_g$ and King $W_0$ for each run. Open circles denote survival to core collapse or 1.5 Gyr and crosses denote disruption. If collapse occurs within 1.5 Gyr, the number above open circles indicate time in units of $10^{10}$ yr at collapse and below the circle, mass in units of initial mass. Otherwise, number below the symbol indicates mass at 1.5 Gyr. The shaded region is an estimate of the clusters that do not survive for a Hubble Time.

Fig. 10.— 70 (non pcc) clusters from Webbink's compilation covering the same range of $R_g$ as in Fig. 9. The disruption boundary is also shown (solid curve).

Fig. 11.— Ratio of cluster edge to dynamically inferred tidal radius for King $W_0 = 7$ clusters and $R_g = 3, 5, 8, 12 \,\mathrm{kpc}$ (labeled).

Fig. 12.— King $W_0 = 7$ at $R_g = 12 \,\mathrm{kpc}$, intial mass function power-law index $x = 3$, and no shocking. Projections of contours of constant power-law index $x$ are shown on the base with the key to the line types shown at left.

Fig. 13.— Same as in Fig. 12 with shocking.